\pdfoutput=1
\documentclass{amia}
\usepackage{graphicx}
\usepackage[labelfont=bf]{caption}
\usepackage[superscript,nomove]{cite}
\usepackage{color}
\usepackage{url}
\let\OLDthebibliography\thebibliography
\renewcommand\thebibliography[1]{
  \OLDthebibliography{#1}
  \setlength{\parskip}{0pt}
  \setlength{\itemsep}{0pt plus 0.3ex}
}

\begin{document}

\title{Focused Clinical Query Understanding and Retrieval of Medical Snippets powered through a Healthcare Knowledge Graph}

\author{Maulik R. Kamdar, Ph.D., Michael Carroll, B.S., Will Dowling, Ph.D., Linda Wogulis, B.S., Cailey Fitzgerald, M.S.,  Matt Corkum, M.S., Danielle Walsh, M.S., David Conrad, B.S., Craig E. Stanley, Jr., Ph.D., Steve Ross, M.D., Dru Henke, B.S.,  Mevan Samarasinghe, M.S.}

\institutes{
    Elsevier, Health and Commercial Markets, Philadelphia, PA\\
}

\maketitle

\section{Introduction}
For the diagnosis, prognosis, and treatment of their patients, clinicians need access to accurate, succinct, updated, and trustworthy information, provided by renowned medical organizations and disseminated through patient guidelines, medical textbooks and journals, and synoptic overviews. Whereas, information systems such as UpToDate\cite{uptodate} and Dynamed\cite{dynamed} provide useful synoptic content in general medicine, synoptic content is often manually curated by teams of domain experts and does not meet the need for specialists. Medical specialists must search and synthesize information on focused, yet esoteric, questions from a broad set of literature sources (textbooks, guidelines, journal articles) in the course of a busy practice using search engines (e.g., ClinicalKey\cite{clinicalkey}, PubMed\cite{pubmedlink}). There are several barriers for clinicians to do focused clinical search at the point of care: growth and evolution of medical knowledge, insufficient time, unanswered questions, additional search on patient comorbidities and contexts, lack of awareness of which resource to search for specialty questions, and skepticism and lack of trust regarding the quality of search results\cite{del2014clinical,cook2013barriers}. 

As highlighted in the Ely taxonomy\cite{Ely429}, the 3 most common categories of questions that clinicians need answered when using computational systems were: \emph{i)} \textit{``What is the drug of choice for condition x?''}, \emph{ii)} \textit{``What is the cause of symptom x?''}, and \emph{iii)} \textit{``What test is indicated in situation x?''}. Advanced technological solutions that automate the search and retrieval of the right snippets from a corpus of trusted medical literature sources in the right context for such questions are critical for the practice of medicine and patient care. Knowledge Graphs can be effective tools to address multiple search and knowledge inference problems in healthcare and biomedicine\cite{kamdar2019enabling}. We present our research and development on a Focused Clinical Search Service (HGFCSS), powered by the Elsevier Healthcare Knowledge Graph (termed HG henceforth), that infers the intent behind focused clinical search queries and retrieves relevant, updated, and trusted medical content from a diverse corpus of synoptic and reference medical literature sources.

\section{Methods}

\subsection{Elsevier Healthcare Knowledge Graph}
Elsevier Healthcare Knowledge Graph\cite{kamdar2020text} (HG) is a platform built to enable advanced clinical decision support and enhanced content discovery. HG includes knowledge and data from heterogeneous healthcare sources about diseases, drugs, findings, guidelines, cohorts, journals, and books. As of August 2020, HG consists of more than 400,000  medical concepts (e.g., drugs, diseases, findings), 1.5 million medical term labels and synonyms for these concepts, and more than 5 million hierarchical and semantic relations (e.g., relations of type \textit{has differential diagnosis}) between these concepts. Subject matter experts (SMEs) curate HG using novel exploration interfaces to keep the medical knowledge regularly updated. Snippets from synoptic and reference medical content are tagged with HG concepts and relations by SMEs or extracted by natural language processing (NLP) models \cite{kamdar2020text}, and are ingested into HG daily through automated pipelines. In this research, a medical snippet comprises the title of the content (e.g., book title, chapter title), section titles, navigational information or `breadcrumbs' for a given section in the content, and sentences. 

\subsection{Focused Clinical Query Understanding}
In this research, we define a clinical search query as a phrase that is provided by clinicians for which he/she wishes to retrieve relevant and precise information from trusted medical literature (e.g., \textit{asthma differential diagnosis}, \textit{temozolomide adverse reactions}, \textit{COVID remdesivir dosage}). Given a clinical search query, the HGFCSS parses and identifies the set of medical concepts and their semantic types from HG, as well as relevant medical phrases (e.g., treatment, diagnosis) or cohorts (e.g., pregnancy), in that query. The parser intelligently infers the clinical query intent and performs corrections on misspelled words. Medical phrases are further interpreted with respect to structural elements in literature sources and HG relation types. The parser also performs query expansion to identify similar concepts in the HG hierarchy and query relaxation to drop irrelevant concepts. Finally, the natural language query is rewritten to a more structured representation including the identified concepts and relations types.

\subsection{Retrieval of Medical Snippets from Literature Sources}
Two automated models (conventional NLP and biomedical word embeddings\cite{kamdar2020text} respectively) were used to tag a diverse set of literature sources ($\geq$ 5,000 documents), of reference textbooks, synoptic curated content, and drug monographs, with HG relations. We have indexed medical content in HG and the HG-tagged literature corpus through multiple approaches. We have developed a federated querying infrastructure over these indexes, so that we can retrieve the right section in the right literature source for focused clinical search queries with a desirable query response time. Using automation methods and regular update of concepts, labels, and relations from several sources, we can ensure that clinicians are always equipped to find recent, succinct, and trustworthy medical content for their focused queries. 

We have evaluated the different methods in the HGFCSS through multiple approaches. Moreover, we have evaluated the HGFCSS against the popular ClinicalKey Search Engine\cite{clinicalkey}, to retrieve the exact and desirable results (i.e., right section in a document) for a gold set of 100 real-world focused clinical search queries over the same corpus, using an average nDCG (Normalized Discounted Cumulative Gain) score computed on the top 10 responses.

\section{Results}
The biomedical word embedding-based model, which was previously evaluated by four experts, had a precision of 61.4\% and recall of 86.3\% for the task of tagging HG relations over a limited corpus\cite{kamdar2020text}. The conventional NLP and biomedical word embedding models were evaluated for tagging HG relations in another limited corpus manually tagged by SMEs. The models had a median coverage of 0.58 and 0.82 respectively with respect to manual tagging. The HGFCSS had an average nDCG score of 0.56 over the ClinicalKey Search Engine (nDCG of 0.38) for the task to retrieve the exact results for a set of focused clinical search queries. The query response time for HGFCSS was analyzed over multiple simulations (300--900 requests per minute) with  99\% queries taking $\leq$ 200ms for retrieval.

\section{Discussion}
Our preliminary evaluation has shown the effectiveness of the novel HGFCSS to retrieve relevant content snippets for clinical search queries. We showcase that structured representation of medical content using knowledge graphs and advanced information retrieval and natural language processing methods can aid in the development of advanced methods for focused clinical search, where the query intent of clinicians is interpreted intelligently and accurate, succinct, updated, and trusted, medical content is retrieved at the point of care. We are planning to extend the HGFCSS’s query parsing framework to use the biomedical word embeddings and learning to rank models to improve the query intent interpretation, to retrieve more related content, and to improve ranking of the search results.

\makeatletter
\renewcommand{\@biblabel}[1]{\hfill #1.}
\makeatother

\bibliographystyle{unsrt}
\bibliography{references}

\begin{thebibliography}{1}

\bibitem{uptodate}
{Wolters Kluwer Health}.
\newblock {UpToDate}.
\newblock \url{https://www.uptodate.com/home}.

\bibitem{dynamed}
{EBSCO Health}.
\newblock {Dynamed}.
\newblock \url{https://www.dynamed.com}.

\bibitem{clinicalkey}
{Elsevier}.
\newblock {ClinicalKey Search Engine}.
\newblock \url{https://www.elsevier.com/solutions/clinicalkey}.

\bibitem{pubmedlink}
\uppercase{US} National Libraries~of Medicine.
\newblock {PubMed}.
\newblock \url{https://www.ncbi.nlm.nih.gov/pubmed/}.

\bibitem{del2014clinical}
Guilherme Del~Fiol et~al.
\newblock Clinical questions raised by clinicians at the point of care: a
  systematic review.
\newblock {\em JAMA internal medicine}, 174(5):710--718, 2014.

\bibitem{cook2013barriers}
David~A Cook et~al.
\newblock Barriers and decisions when answering clinical questions at the point
  of care: a grounded theory study.
\newblock {\em JAMA internal medicine}, 173(21):1962--1969, 2013.

\bibitem{Ely429}
John~W Ely et~al.
\newblock A taxonomy of generic clinical questions: classification study.
\newblock {\em BMJ}, 321(7258):429--432, 2000.

\bibitem{kamdar2019enabling}
Maulik~R Kamdar et~al.
\newblock Enabling web-scale data integration in biomedicine through linked
  open data.
\newblock {\em NPJ digital medicine}, 2(1):1--14, 2019.

\bibitem{kamdar2020text}
Maulik~R Kamdar et~al.
\newblock Text snippets to corroborate medical relations: An unsupervised
  approach using a knowledge graph and embeddings.
\newblock {\em AMIA Summits on Translational Science Proceedings}, 2020:288,
  2020.

\end{thebibliography}

\end{document}